\documentclass[trackchanges]{aastex701}

\usepackage{amsmath}
\usepackage{subcaption}
\usepackage{caption}
\usepackage{longtable}
\begin{document}

\title{Multi-Wavelength Afterglows as Diagnostic Probes of Dense Circumburst Medium in GRBs}

\author[0000-0003-3659-4800]{Xiao-Hong Zhao}
\affiliation{Yunnan Observatories, Chinese Academy of Sciences, Kunming, China; zhaoxh@ynao.ac.cn}
\affiliation{Center for Astronomical Mega-Science, Chinese Academy of Sciences, Beijing, China}
\email{zhaoxh@ynao.ac.cn}

\begin{abstract}
Gamma-ray bursts (GRBs) are generally believed to occur in environments where the surrounding medium is either a uniform interstellar medium (ISM) or, in some cases, a dense stellar wind from a massive progenitor. Recently, GRB 191019A has been proposed to originate within the accretion disk of an active galactic nucleus (AGN), suggesting that some GRBs may occur in extremely dense environments, although this interpretation remains under debate. 
This scenario has drawn considerable attention, as AGN disks are promising sites that can host progenitors of both long and short GRBs, and whose dense, gas-rich environment could significantly influence jet propagation and afterglow emission. Yet, our theoretical understanding of the resulting afterglow signatures in such environments is limited, and further systematic exploration is required. In this study, we investigate how multi-wavelength afterglow light curves can be utilized as diagnostic tools to probe the nature of the circumburst environment. Our results show that in dense environments, GRB afterglows exhibit distinct frequency-dependent behaviors. For jets with large opening angles, the X-ray light curve displays a shallow decay or bump due to a transition from synchrotron to SSC dominance, while the optical and high-energy (GeV) light curves follow typical power-law decays. On the other hand, for small opening angles, the light curves exhibit wavelength-dependent jet breaks: the GeV and optical bands break simultaneously, while the X-ray break is delayed as the SSC component gradually compensates for the fading synchrotron component. These signatures provide potential diagnostics of GRBs occurring in dense media such as AGN disks.

\end{abstract}

\keywords{Active galactic nuclei (16) ---Gamma-ray bursts (629) ---Relativistic jets (1390) }

\section{Introduction} \label{sec:intro}

Gamma-ray bursts (GRBs) are among the most energetic phenomena in the universe, and their afterglow emission provides critical insights into the nature of the relativistic shock and the circumburst environment. The standard afterglow model attributes the observed emission to synchrotron radiation from relativistic electrons accelerated in the external forward shock as the GRB jet runs into the circumburst medium \citep{1997ApJ...476..232M,1998ApJ...497L..17S}. Based on multi-wavelength afterglow fitting, the circumburst medium of both long bursts and short bursts is found generally to be consistent with a uniform interstellar medium (ISM) or intergalactic medium (IGM), while only a small fraction of bursts require a dense stellar wind profile \citep{2002ApJ...571..779P,2003ApJ...597..459Y,2011A&A...526A..23S,2015ApJ...815..102F}.

However, some bursts were believed to occur in much denser environments, such as dense molecular clouds \citep{daiAfterglowGRB9901231999,lazzatiIronLineAfterglow1999}. Recent observational developments further suggest that some GRBs may occur in environments far denser than the typical ISM, like the accretion disks of active galactic nuclei (AGNs). A possible example is GRB 191019A. GRB 191019A is classified as a long-duration burst with a redshift $z=0.248$, yet no associated supernova has been identified \citep{2023NatAs...7..976L}. The burst’s host galaxy is dominated by an old stellar population, which is atypical for long GRBs that are usually linked to star-forming galaxies (e.g., \citealt{fruchterLongGrayBursts2006}). Moreover, the burst occurred near the center of its host galaxy \citep{2023NatAs...7..976L}. These properties have led to the speculation that GRB 191019A may be an intrinsically short burst, originating from a compact binary neutron star merger within a dense accretion disk and the observed long duration was attributed to the effect of the dense circumburst environment (\citealt{lazzatiPromptEmissionGRay2022,lazzatiGRB191019AShort2023a}, but see \citealt{strattaPuzzlingLongGRB2025}).

In theory, the AGN disks are considered to be potential sites for both in situ star formation and the dynamical capture of stars from surrounding nuclear star clusters \citep{1993ApJ...409..592A,2003MNRAS.339..937G,2020MNRAS.499.2608F,2024MNRAS.528.4958W}.
As a result, AGN disks may host massive stars and compact object binaries. The core collapse of massive stars formed or captured in these disks, or the mergers of neutron star–neutron star (NS–NS) or neutron star–black hole (NS–BH) binaries embedded within them, can produce long and short GRBs, respectively \citep{2020MNRAS.498.4088M,2020ApJ...898...25T,pernaAccretionInducedCollapseNeutron2021}. Recent gravitational wave detections provide additional support for this scenario. The presence of objects within unexpected mass ranges, such as a component in the pair-instability mass gap ($\sim$64–135 $M_\odot$, e.g., GW190521, \citealt{ligoscientificcollaborationandvirgocollaborationGW190521BinaryBlack2020}) and a component in the lower mass gap ($\sim$2.2–5 $M_\odot$, e.g., GW190814, \citealt{2020ApJ...899L...1Z}), poses challenges to standard stellar evolution models, but can be naturally explained if the mergers occurred in AGN disks \citep{yangBlackHoleFormation2020,tagawaEccentricBlackHole2021,tagawaMassgapMergersActive2021}. Therefore, GRB occurrences within dense AGN disks represent a plausible theoretical expectation. 

If GRBs occur in such dense environments, the jet propagation dynamics and corresponding afterglow signatures are significantly different from the classical scenario of a low-density ISM. Some studies argue that GRB jets can successfully emerge from AGN disks under specific conditions (e.g., around lower-mass supermassive black holes) and produce non-standard afterglows with unique features like density-dependent light curves with an earlier peak, while others suggest that jets with lower power may be choked within denser regions, failing to generate canonical afterglow signatures but instead emitting detectable shock breakout emission and subsequent shock-cooling radiation (e.g., \citealt{pernaElectromagneticSignaturesRelativistic2021,2021ApJ...911L..19Z,renInteractingKilonovaeLonglasting2022,yuanMultibandEmissionTwocomponent2024a,zhangPropagationGammaRayBurst2024a,chenObservationalPropertiesThermal2025}). The afterglow emission in dense environment could be absorbed or scattered. Recent efforts have examined the radiative transfer or radiative diffusion of GRB afterglow emission through dense AGN disk material, showing that synchrotron self-absorption (SSA) significantly suppresses radio emission, optical emission and even X-ray emission  \citep{2022MNRAS.516.5935W,2023MNRAS.521.4233R,2024ApJ...972..101K,2025OJAp....8E..23K}.

The choked-jet scenario discussed above requires GRBs to occur in the extremely dense regions (with the density $n>10^{10}$ cm $^{3}$) of AGN disks. However, the density of AGN disks can vary by many orders of magnitude, from $\sim10^3$ to $10^{15} {\rm cm}^{-3}$, depending on the radial location within the disk and the mass of the central black hole (e.g., \citealt{2020MNRAS.499.2608F}). This wide range of densities introduces significant variation in the expected afterglow signatures. In very dense regions, radiative transfer effects can severely distort or even obscure the afterglow emission. In this work, we focus on the intermediate-density regime of $n \sim 10^3$–$10^6 {\rm cm}^{-3}$, corresponding to the outer regions of AGN disks (assuming the GRB jet propagates perpendicular to the AGN disk for simplicity). In these environments, SSA suppression is relatively weak, but the afterglow light curves differ significantly from those in the typical ISM case.

In such intermediately dense environments, the inverse Compton (IC) processes, particularly the synchrotron self-Compton (SSC) emission, can become dominant at early times (within thousands of seconds after the GRB trigger), especially in the X-ray band. The transition from synchrotron-dominated to SSC-dominated emission may produce distinct features in the light curves. 
Our goal is to investigate how dense environments shape the afterglow emission across wavelengths and whether these features can serve as observational diagnostics for identifying the GRB's environment.

The paper is organized as follows. In Section 2, we present the analytical evolution of the characteristic synchrotron frequencies in dense environments. Section 3 provides the analytical afterglow light curves across different bands. In Section 4, we examine the $\gamma\gamma$ absorption of high-energy photons and show that, due to the high circumburst densities, the SSC emission is expected to be significantly attenuated above $\sim$ 10 GeV. Section 5 provides numerical light curve calculations across a range of model parameters, revealing characteristic features of GRB afterglows in dense environments. In Section 6, we discuss the implications of our results and summarize our conclusions.

\section{critical frequencies in a dense circumburst environment}
In the standard GRB afterglow model of GRBs, the afterglow emission arises from a spherical shock, whose spectrum is generally divided into four distinct power-law segments by three critical frequencies \citep{1998ApJ...497L..17S}: the synchrotron typical frequency $\nu_m$, the cooling frequency $\nu_c$, and the SSA frequency $\nu_a$. The spectral slopes in each segment depend on the ordering of these three frequencies. The afterglow light curves thus also depend on the evolution of the three frequencies and the overall normalization.  For a typical ISM medium of $n\sim$ several,  a given isotropic-equivalent forward shock energy $E_{\rm iso}$, and an observed time $T$, $\nu_m$ and $\nu_c$ can be written as \citep{1998ApJ...497L..17S}
\begin{eqnarray}
     \nu_m&=&2.5 \times 10^{13}\Big(\frac{z_1}{1.3}\Big)^{1/2}E_{\rm iso, 51}^{1/2}\epsilon_{B,-3}^{1/2}\epsilon_{e,-1}^{2}T_3^{-3/2}\text{Hz}\\
    \nu_c&=&2.2 \times 10^{18}\Big(\frac{z_1}{1.3}\Big)^{-1/2}E_{\rm iso,51}^{-1/2}\epsilon_{B,-3}^{-3/2}n_0^{-1}(1+Y)^{-2}T_3^{-1/2}\text{Hz}.
\end{eqnarray}
where $z$ is the redshift (here z=0.3 is used), $z_1=1+z)$, Y is the Compton Y factor \citep{sariSynchrotronSelfComptonEmission2001}, given by $Y= \frac{-1+\sqrt{1+\frac{4\eta\epsilon_e}{\epsilon_B}}} {2}$, and $\eta$ is the radiation efficiency can be estimated as $\eta = 1$ for fast cooling and $\left(\frac{\gamma_m}{\gamma_c}\right)^{p-2}$ for slow cooling. The redshift of $z = 0.3$ is chosen because GRBs occurring in dense environments tend to exhibit rapidly decaying afterglows with earlier jet breaks, making it difficult to obtain well-sampled multi-band light curves. Therefore, we focus on low-redshift bursts, which are intrinsically brighter and more readily detected across multiple wavelength bands, making it easier to confirm whether they occur within an AGN disk. The recent burst GRB 191019A was believed to occur within an AGN disk at a redshift of $z = 0.248$ \citep{2023NatAs...7..976L}, demonstrating that such nearby bursts can be observed. We thus adopt a similar redshift in our paper. Note that we use the notation $Q=10^xQ_x$ in cgs units throughout the paper. 

Electrons in the afterglow shock transit from the fast cooling regime to the slow cooling regime with the shock evolution. With the typical parameters, the transition time is $T_{\rm tran}= 1.3(\frac{z_1}{1.3})E_{\rm iso, 51}\epsilon_{B,-3}^{2}\epsilon_{e,-1}^{2}n_0(\frac{1+Y}{10.5})^{2}~\text{s}$. Note that when $\epsilon_e = 0.1$ and $\epsilon_B=0.001$, $1+Y\approx 10.5$ for fast cooling. However, in a dense environment, the afterglow emission can remain in the fast cooling regime for over 10 days, i.e.,
\begin{eqnarray}
    T_{\rm tran}= 14.8(\frac{z_1}{1.3})E_{\rm iso, 51}\epsilon_{B,-3}^{2}\epsilon_{e,-1}^{2}n_6\Big(\frac{1+Y}{10.5}\Big)^{2}~\text{days}
\end{eqnarray}
In an extremely dense environment in AGN disks, the synchrotron cooling Lorentz factor $\gamma_c^{\rm syn}$ could be less than unity,  i.e.,  
$\gamma_c^{\rm syn} \approx 0.4 \left( \frac{z_1}{1.3} \right)^{-1/8} E_{\text{iso},51}^{-3/8} \epsilon_{B,-3}^{-1} n_{10}^{-5/8} T^{1/8}_3$,
rendering it physically not the cooling Lorentz factor again. The IC cooling further makes the electrons cool down to lower energies due to $\gamma_c=\gamma_c^{\rm syn}/(1+Y)$. This implies that the electrons would cool down to a non-relativistic speed within the dynamic time scale if no other processes worked.  However, in this situation, the SSA can prevent the electrons from cooling down to below an energy of $\gamma_a= \gamma_a(\nu_a)>\gamma_c$, where $\nu_a$ is the SSA frequency defined by the SSA opacity $\tau_a(\nu_a)=1$. Thus $\gamma_c$ is unimportant since the electrons lie in the optical thick region.  While synchrotron cooling in this case is inefficient, electrons below $\gamma_a$ can still lose energy through IC scattering. This occurs because synchrotron photons can be up-scattered to frequencies higher than the SSA frequency, carrying away the electrons' energy. Thus electrons can further cool down to a lower energy \citep{ghiselliniThermalizationSynchrotronAbsorption1998}.

For $\gamma_c<1$, the usual treatment that the flux is normalized with the peak flux at $\nu_c$ is not appropriate because the synchrotron radiation at $\nu_c$ no longer works. We can use the flux at $\nu_m$ as the normalization. 
In this case, the comoving electron distribution per unit volume can be given by
\begin{equation}
    n'_{\gamma_e}=n'_{\gamma_{e0}}\begin{cases}
        \Big(\frac{\gamma_e}{\gamma_m}\Big)^{-2} & \gamma_e<\gamma_m \\
      \Big(\frac{\gamma_e}{\gamma_m}\Big)^{-p-1} &\gamma_m<\gamma_e<\gamma_{\text{max}},       
    \end{cases}
\end{equation}
where $\gamma_m$, $\gamma_{\rm max}$, and $p$ are the minimum Lorentz factor, the maximum Lorentz, and the power-law index of the shocked electrons.
The normalization is $n'_{\gamma_{e0}} = 4 \gamma n_{\gamma_{e0}} =n'_et_mz_1/(\gamma_m\gamma T)$, where $t_m=\frac{6\pi m_ec}{\sigma_TB^2\gamma_m(1+Y)}$ is the comoving time within which high energy electrons cool down to $\gamma_m$ and $\gamma$ is the bulk Lorentz factor. Thus the observed flux at $\nu_m$ can be given by
$F_{\nu_m}= \frac{z_1}{4\pi D_L^2} n_{\gamma_{e0}}\gamma_m \frac{4\pi}{3} R^3P_\nu $, where $D_L$ is the luminosity distance and $R=4 \gamma^2 c T$ is the radius of the shock. The SSA frequency can be given by different expressions, depending on the specific ordering of the three critical frequencies. The absorption coefficient of SSA is given by (e.g., \citealt{panaitescuAnalyticLightCurves2000,wuOpticalFlashesVery2003})
\begin{eqnarray}
    \alpha'_{\nu'} =
    \begin{cases}   c_{1}q_eB^{-1}n'_{\gamma_{e0}}\gamma_m^2\gamma_c^{-6}\nu_c^{5/3}\nu_a^{-5/3} & \nu_a<\nu_c<\nu_m\\
        c_{2}(2\pi m_ec)^{-3}\gamma^3 q_e^4B^2n'_{\gamma_{e0}}\gamma_m^2 \nu_a^{-3}& \nu_c<\nu_a<\nu_m\\
        c_3 q_e B_m^{-1}n'_{\gamma_{e0}}\gamma_m^{-4} \nu_m^{(p+5)/2}\nu_a^{-(p+5)/2}   & \nu_c<\nu_m<\nu_a.
    \end{cases}
\end{eqnarray} 
Using $\delta R'\approx R/12\gamma$ and $\tau'_{a} =\alpha'_{\nu'}\delta R'=1$, we can find the observed SSA frequency is given by
\begin{eqnarray} \label{nua_fastcool}
    \nu_{a} =
    \begin{cases}  
   4.0\times 10^{14} (\frac{z_1}{1.3})^{-1/2}E_{\rm iso, 51}^{7/10}\epsilon_{B,-3}^{6/5}n_{6}^{11/10}\frac{1+Y}{10.5}T_3^{-1/2}\text{Hz} & \nu_a<\nu_c<\nu_m \\
6.4\times 10^{12} (\frac{z_1}{1.3})^{1/2}E_{\rm iso, 51}^{1/6}n_{6}^{1/6}(\frac{1+Y}{10.5})^{-1/3}T_3^{-1/2}\text{Hz} & \nu_c<\nu_a<\nu_m\\
       8.5\times 10^{12} (\frac{z_1}{1.3})^{2/(p+5)}E_{\rm iso, 51}^{(p+1)/2(p+5)}\epsilon_{e,-1}^{2(p-1)/(p+5)}\epsilon_{B,-3}^{(p-1)/2(p+5)}\\
       ~~~~~~ \times n_{6}^{1/(p+5)}(\frac{1+Y}{10.5})^{-2/(p+5)}T_3^{-3(p+1)/2(p+5)}\text{Hz} & \nu_m<\nu_a.
    \end{cases}
\end{eqnarray} 
For not very dense circumburst, the slow cooling can be relevant. In this case, if the SSC cooling is the dominated cooling mechanism, the cooling Lorentz factor is $\gamma_c\approx \left(\frac{\epsilon_B}{\epsilon_e}\right)^{\frac{1}{4-p}}  \left(\gamma_{\rm c}^{\rm syn}\right)^{\frac{2}{4-p}} \gamma_m^{\frac{2-p}{4-p}}$. The SSA frequency in this case is given by (e.g., \citealt{wijersPhysicalParametersGRB1999,panaitescuAnalyticLightCurves2000,wuOpticalFlashesVery2003})
\begin{eqnarray}\label{nua_slowcool}
    \nu_{a} =
    \begin{cases}  
   1.3\times 10^{13} (\frac{z_1}{1.3})^{-1} E_{\rm iso, 51}^{1/5}\epsilon_{B,-3}^{1/5}\epsilon_{e,-1}^{-1}n_{6}^{3/5}\text{Hz} & \nu_a<\nu_m<\nu_c \\
1.8\times 10^{13} (\frac{z_1}{1.3})^{(p-6)/2(p+4)}E_{\rm iso, 51}^{(p+2)/2(p+4)}\epsilon_{B,-3}^{(p+2)/2(p+4)}\epsilon_{e,-1}^{2(p-1)/(p+4)}\\
~~~~\times n_{6}^{2/(p+4)}T_3^{-(3p+2)/2(p+4)}\text{Hz} & \nu_m<\nu_a<\nu_c.
    \end{cases}
\end{eqnarray} 
The case of $\nu_a>\nu_c>\nu_m$ is not relevant.

\begin{figure}[htbp!]
      \centering
        \centering
    \includegraphics[width=\textwidth]{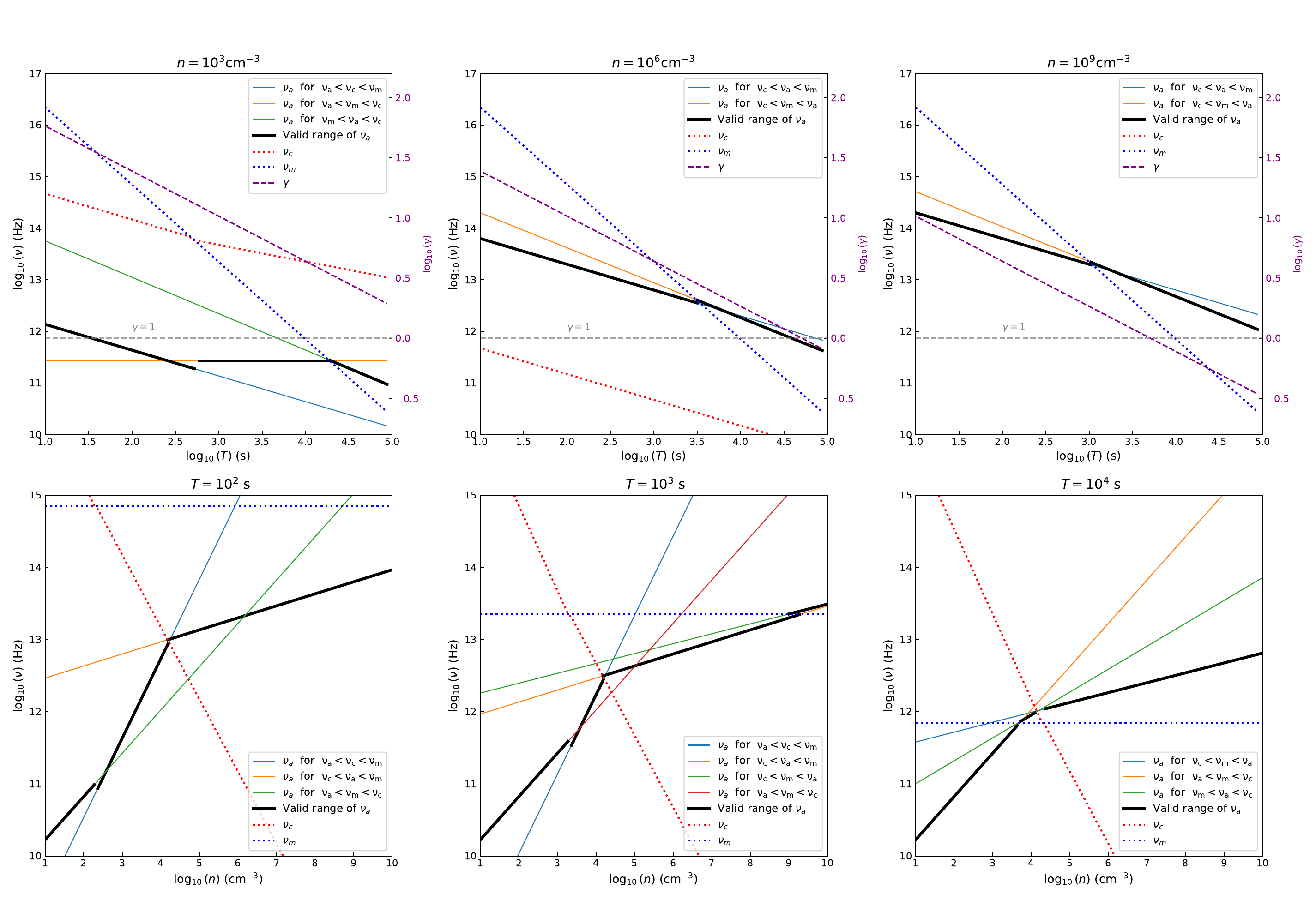}
    
    \caption{Temporal evolution and density dependence of the synchrotron critical frequencies. Upper row: frequency evolution with time observed $T$ for different ambient densities. The secondary y-axis on the right shows the evolution of the bulk Lorentz factor $\gamma$, indicated by purple dashed lines. Note that the analytical evolution of the synchrotron critical frequencies is only valid for $\gamma\gg1$ and breaks down when $\gamma$ approaches unity. Lower row: frequency dependence of density n at different times. The adopted  parameters include $E_{\rm iso}=1\times 10^{51}$ erg, $\epsilon_B=0.001$, $\epsilon_e=0.1$, $p=2.3$.} 
    \label{evolu_nu}
\end{figure}

Fig. \ref{evolu_nu} shows the dependence of the three synchrotron critical frequencies on the observed time and the circumburst density using some typical parameters: $E_{\rm iso}=1\times 10^{51}$ erg, $\epsilon_B=0.001$, $\epsilon_e=0.1$, $p=2.3$. In very early afterglow phases ($\sim$ hundreds of seconds), the afterglow emission can include complex components, including the reverse shock emission and the steep decay phase \citep{sariPredictionsVeryEarly1999,zhangPhysicalProcessesShaping2006}. To consider the light curves under the standard afterglow model, we mainly consider afterglow emission after T=100 s. As shown in the figure, the SSA frequency $\nu_a$ falls into the regime of $\nu_c<\nu_a<\nu_m$ or $\nu_c<\nu_m<\nu_a$ across most of the parameter space in dense environments ($n\geq 10^6$ cm$^{-3}$) and generally remains below the optical band ($\sim 10^{14}–10^{15}$ Hz). Some papers claimed that the SSA frequency can be as high as optical, even X-ray bands (e.g., \citealt{2022MNRAS.516.5935W}). This might be plausible in very early times or an extremely dense environment $n\gg$ 10$^{10}$ cm$^{-3}$. We note that only in the regime of $\nu_a<\nu_c<\nu_m$, $\nu_a$ is sensitive to $n$, scaling as $\nu_a\propto n^{11/10}$ but in other regimes the dependence is very weakly (see eq.\ref{nua_fastcool} and eq. \ref{nua_slowcool}), making $\nu_a$ hard to exceed the optical band. For a moderately dense environment ($n<100 \rm cm^{-3}$), the afterglow emission has transitioned from the fast cooling to the slow cooling regime by $\sim 1000$ s (see the lower panels in Fig. \ref{evolu_nu}). In contrast, for dense environments ($n>10^6$ cm$^{-3}$), the emission remains in the fast cooling regime for longer than 10$^5$s. In extreme dense region (e.g., $n=10^9$ cm$^{-3}$), the shock rapidly decelerates, reaching $\gamma\sim1$ as early as $T \sim 3000$s.

\section{analytical light curves} \label{analytical_lc}
\begin{figure}[htbp!]
    \centering
        \centering
    \includegraphics[width=\textwidth]{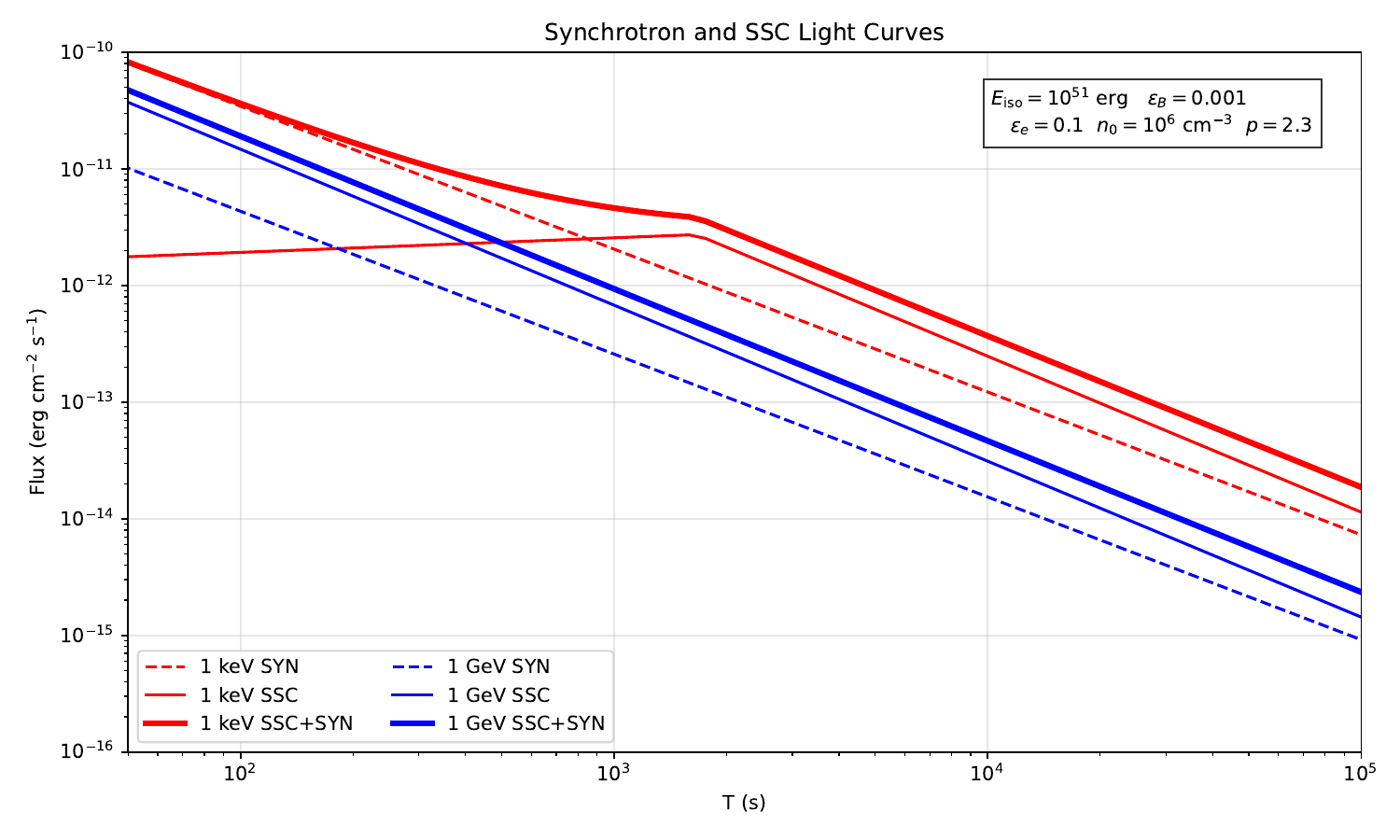}

    \caption{Analytical afterglow light curves in the X-ray and GeV bands, including the synchrotron component and the SSC component. } 
    \label{lc_analy}
\end{figure}

The evolution of four critical frequencies determines the afterglow light curves. The analytical afterglow light curves in the ISM and wind medium have been given \citep{1998ApJ...497L..17S,panaitescuAnalyticLightCurves2000}. The contribution of the SSC emission from the afterglow to high-energy GeV emission has also been considered (e.g., \citealt{panaitescuAnalyticLightCurves2000,dermerBeamingBaryonLoading2000,zhangGammaRayBurstAfterglow2001}). In dense and homogeneous media, the X-ray or higher-energy light curves include the synchrotron component and the IC component and are mainly determined by two critical frequencies: $\nu_m$ and the corresponding SSC frequency $\nu^{\rm SSC}_m=\gamma_m^2\nu_m$, which are given by
\begin{align}\label{nu_m_num_ssc_T}
\nu_m &\approx 2.2\times 10^{13} \, E_{\text{iso},51}^{1/2} \epsilon_{B,-3}^{1/2} \epsilon_{e,-1}^2 T_3^{-3/2}~  \rm Hz\\
\nu_{m}^{\rm SSC}&\approx 7.6\times 10^{17} \, E_{\text{iso},51}^{3/4} \, \epsilon_{B,-3}^{1/2} \, \epsilon_{e,-1}^4 \, n_{6}^{-1/4} \, T_3^{-9/4}  \rm Hz.
\end{align}
The flux density ($F_\nu$) is for the synchrotron case and SSC case are respectively
\begin{align}
F_{\nu} \approx 37.2
\begin{cases}\label{Fnu_T_syn}
\left(\frac{z_1}{1.3}\right)^{-1}d_{L,27.7}^{-2}E_{\text{iso},51}^{3/4}  \epsilon_{B,-3}^{-1/4} \left(\frac{1+Y}{10.5}\right)^{-1}\nu_{13.3}^{-1/2}T_3^{-1/4}   \rm mJy & \nu<\nu_m,\\
 \left(\frac{z_1}{1.3}\right)^{-1}d_{L,27.7}^{-2}E_{\text{iso},51}^{(2+p)/4} \epsilon_{B,-3}^{(p-2)/4} \epsilon_{e,-1}^{p-1} \left(\frac{1+Y}{10.5}\right)^{-1}\nu_{13.3}^{-p/2} T_3^{(2-3p)/4}    ~ \rm mJy & \nu>\nu_m
\end{cases}
\end{align}
\begin{align}\label{Fnussc_T}
F^{\text{ssc}}_\nu\approx 6.0\times 10^{-4}
\begin{cases}
\left(\frac{z_1}{1.3}\right)^{-1} d_{L,27.7}^{-2} E_{\text{iso},51}^{5/8} n_{6}^{1/8} \epsilon_{B,-3}^{-5/4}  \left(\frac{1+Y}{10.5}\right)^{-2}\nu_{17.9}^{-1/2}T_3^{1/8} \rm mJy  &\nu<\nu_m^{\text{SSC}} \\
 \left(\frac{z_1}{1.3}\right)^{-1} d_{L,27.7}^{-2}E_{\text{iso},51}^{(2+3p)/8} \epsilon_{B,-3}^{(p-6)/4} \epsilon_{e,-1}^{2(p-1)} n_{6}^{(2-p)/8} \left(\frac{1+Y}{10.5}\right)^{-2} \nu_{17.9}^{-p/2} T_3^{(10-9p)/8} \rm mJy   &\nu>\nu_m^{\text{SSC}} 
\end{cases}
\end{align}
It can be seen that in the X-ray band, the temporal behavior of the light curve depends on the dominant radiation mechanism. For the synchrotron-dominated case, the light curve decays with a slope of $(3p-2)/4$. While for the SSC-dominated case, the light curve initially rises with a slope of $1/8$, followed by a decay with a slope of $(10-9p)/8$. This decay slope is slightly steeper than that of the synchrotron-dominated case.
For example, if p = 2.5, the synchrotron decay slope is 1.4, while the SSC decay slope is 1.6. 
Fig. \ref{lc_analy} depicts the analytical light curves. The X-ray light curve initially exhibits a synchrotron-dominated phase (t $\lesssim$ 2000 s), followed by a gradual transition to SSC dominance. This transition leads to a shallow decay segment and eventually a steeper decline. In cases where the SSC component significantly exceeds the synchrotron component, the X-ray light curve may even exhibit a rising phase before entering the decay stage. On the other hand, we also indicate the high-energy (GeV) light curve in the figure, which is dominated by the SSC component from the very beginning and remains a power-law decline. The light curves diverge significantly between bands, reflecting distinct emission regimes.  
The critical time after which the SSC component exceeds the synchrotron component in the X-ray band is 
\begin{eqnarray}\label{tcr}
    T_{\text{cr}}\approx 850 E_{\text{iso},51}^{1/3} \,
\epsilon_{B,-3}^{\frac{2(3 + p)}{6p - 3}} \,
\epsilon_{e,-1}^{\frac{8(p - 1)}{6p - 3}} \,
n_{6}^{\frac{1}{3 - 6p}} \,
\nu_{\text{1keV}}^{\frac{4(1 - p)}{6p - 3}} \,
\left(\frac{1+Y}{10.5}\right)^{\frac{8}{6p - 3}}s.
\end{eqnarray} 
We can find the critical time
to be on the order of thousands of seconds, is relatively sensitive to the microphysical parameters  $\epsilon_B$ and $\epsilon_e$, but weakly dependent on the circumburst medium density or isotropic equivalent energy. For the typical parameters, the critical time is $T_{\rm cr} \sim \mathcal{O}(10^3  \text{s})$, suggesting that if a GRB occurs in a dense environment such as an AGN disk, a shallow decay or a rising phase may appear in the X-ray afterglow light curve around $\sim 1000$ s after the burst.

\section{expected highest photon energy from afterglows in dense environments} \label{maxenergy}
As shown in the last section, in dense environments, the high-energy (GeV) emission can be generated by SSC and synchrotron mechanisms. However, such high-energy emission is subject to strong attenuation from pair production within the afterglow emission region. The efficiency of this attenuation process depends critically on the spectral properties of the lower-energy target photons, which can be either from synchrotron emission or from SSC emission. To determine the total opacity, we must evaluate the contributions from both components. 

\begin{figure}[htbp!]
    \centering
        \centering
    \includegraphics[width=\textwidth]{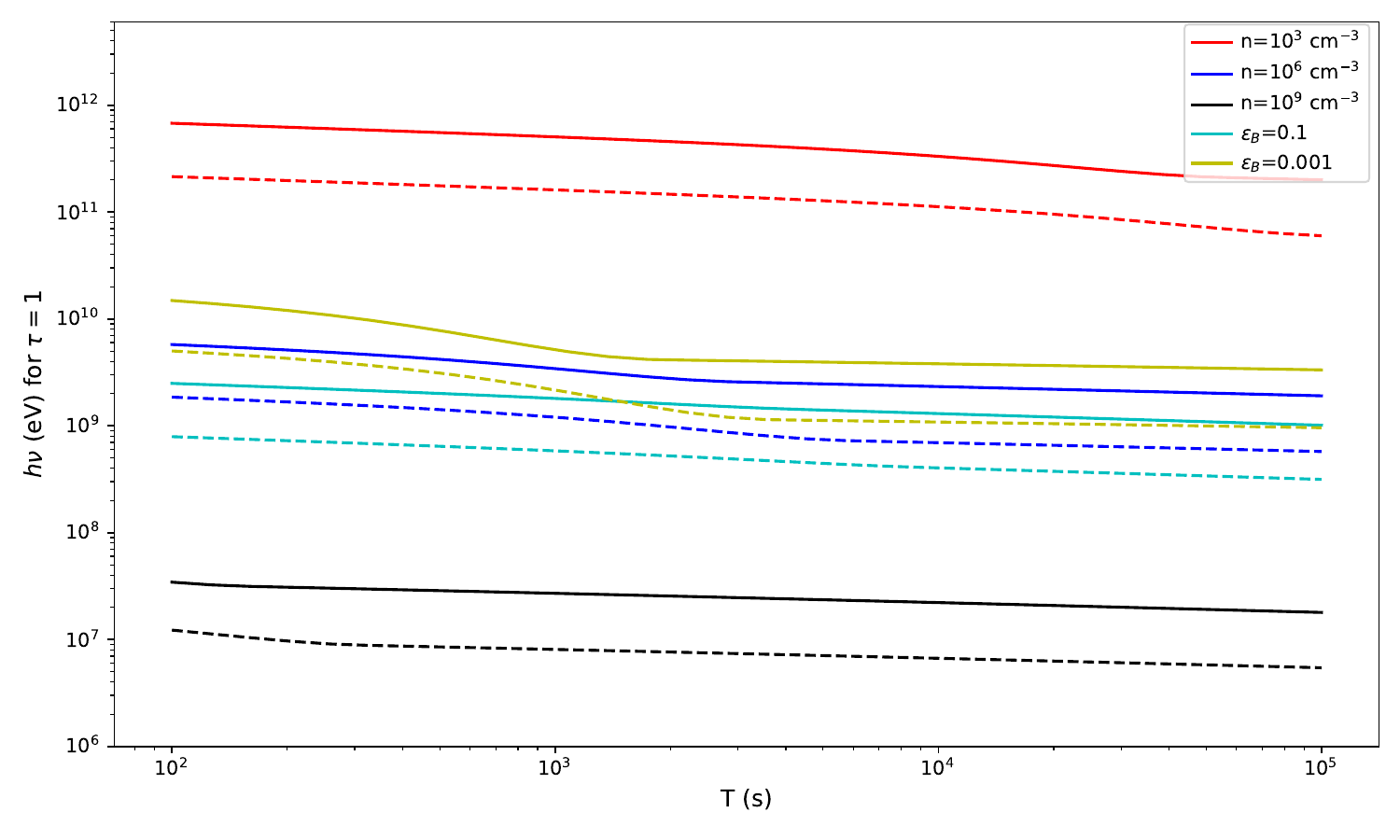}

    \caption{Evolution of the maximum photon energy from the GRB afterlow emission. The solid line and the dashed line are for $E_{\rm iso}$=10$^{51}$ erg and $E_{\rm iso}$=10$^{53}$ erg , respectively. When a parameter varies, the others are fixed at their typical values, as taken in Fig. \ref{evolu_nu}. Note that the break in the curves marks the transition from the synchrotron-dominated to the SSC-dominated regime.} 
    \label{E-T}
\end{figure}

Following Eq. \ref{Fnussc_T}, the SSC spectrum consists of two distinct components: a low-energy part with slope -1/2 and a high-energy part with slope -p/2. Assuming photons with frequency of $\nu$ are annihilated, the photon number of the contributing to the annihilation is $N_{>\nu_{\rm ann}}=\int_{\nu_{\rm ann}}^\infty  F_{\tilde\nu}/(h \tilde\nu) 4\pi D_L^2 t/z_1^2d\tilde\nu$, where $\nu_{\rm ann}=(\gamma m_ec)^2/(z_1^2h\nu) $, and the $\gamma\gamma$ absorption optical depth is  $\tau_{\gamma\gamma}=0.1\sigma_TN_{>\nu_{\rm ann}}/(4\pi R^2)$ (e.g., \citealt{lithwickLowerLimitsLorentz2001,zhangGammaRayBurstAfterglow2001}). As an estimation, here we use $\sigma_{\gamma\gamma}=0.1\sigma_T$ as the annihilation cross section of photon pair. The optical depth for the SSC dominated case is  
\begin{eqnarray}
\tau_{\rm ssc,low} &\approx& 0.6 \, (h\nu)_{{\rm GeV}}^{1/2} \,  
 \epsilon_{B,-3}^{-5/4} n_{6}^{3/4} \,\left(\frac{1+Y}{10.5}\right)^{-2} T_3\\
\tau_{\rm ssc,high} &\approx& 0.2 \, (h\nu)_{{\rm GeV}}^{p/2} \, E_{{\rm iso},51}^{(p - 1)/4} \, \epsilon_{B,-3}^{(p - 6)/4} \, \epsilon_{e,-1}^{2(p - 1)} \, n_{6}^{3/4} \,  \left(\frac{1+Y}{10.5}\right)^{-2}T_3^{(7 - 3p)/4}.
\end{eqnarray}
The synchrotron component may also contribute to emission up to the GeV energy range with the spectrum being a single power-law of $-p/2$. The optical depth for the synchrotron-dominated case is
\begin{eqnarray} \label{tausyn}
\tau_{\rm syn} &\approx& 0.1 \, 
(h\nu)_{{\rm Gev}}^{p/2} \,
E_{{\rm iso},51}^{p/8} \, \epsilon_{B,-3}^{(-2 + p)/4} \, \epsilon_{e,-1}^{p - 1} \, 
 n_{6}^{(4 + p)/8} \, \left(\frac{1+Y}{10.5}\right)^{-2} T_3^{(8 - 3p)/8}.
\end{eqnarray}
The total optical depth is the combined contributions of the synchrotron and the SSC components. Fig. \ref{E-T} shows the evolution of the maximum photon energy (where $\tau(h\nu)=1$) from the GRB afterglow emission. It can be seen that the maximum energy weakly depends on the observed time $T$, the isotropic energy $E_{\rm iso}$, and $\epsilon_B$, but is sensitive to the circumburst density. For example, when $n \lesssim 10^3$ cm$^{-3}$, photons with energies up to $\sim 1$ TeV can escape, whereas in much denser environments ($n > 10^6$ cm$^{-3}$), the high-energy emission is typically limited to below 10 GeV. Therefore, in dense environments like AGN disks, internal $\gamma\gamma$ absorption effectively imposes an upper limit on the observable afterglow photon energy, typically around 10 GeV.

\section{numerical consideration}

In Section \ref{analytical_lc}, we have presented analytical light curves that illustrate the basic scaling relations of synchrotron and SSC emission in dense environments. While these analytical results provide physical intuition, they rely on simplified assumptions  (e.g., such as spherical shock geometry, persistent ultrarelativistic evolution, and simplified treatment of electron distribution evolution). In reality, the afterglow emission is generated from a jet \citep{rhoadsHowTellJet1997,sariJetsGammaRayBursts1999}. The cooling history of non-thermal electrons, involving the synchrotron cooling and SSC cooling, should be considered self-consistently. The invoked assumptions in analytical consideration may introduce deviations from the realistic scenarios. Thus numerical method is used to calculate the dynamic evolution and then the light curves of GRB afterglows.
\subsection{numerical method}

The realistic dynamic evolution of the afterglow shock involving the jet geometry and the sideways expansion has been widely investigated \citep{huangOverallEvolutionJetted2000,granotLateralExpansionGammaray2012a,duffellDecelerationSpreadingRelativistic2018}. The treatment of sideways expansion is non-trivial but important in a dense environment. We employ the dynamic framework developed by \citet{ryanGammaRayBurstAfterglows2020}, which incorporated recent advancements in the lateral expansion treatment and resulted in similar light curves to those from multi-dimensional hydrodynamic simulations \citep{vaneertenGammaRayBurstAfterglow2012}. For simplicity, we consider only a top-hat jet model with a half-opening angle of $\theta_0$, assuming the line of sight lies along the jet axis (see \citealt{zhaoAfterglowLightCurves2022} for further details). No structured-jet configurations are considered.

Instead of adopting the analytical treatment of Ryan et al. (2020) for the electron emission from the forward shock, we first self-consistently compute the electron energy distribution at each radius. This distribution is then used to evaluate the local emission spectrum, and the observed flux is finally obtained by integrating over the equal-arrival-time surface (see \citealt{zhaoAfterglowLightCurves2022} for details). 

The evolution of electron distribution within the jet involves multiple competing energy-loss mechanisms, including adiabatic cooling, synchrotron cooling, and SSC cooling. In the GRB emission region, for a high-energy electron with energy of $\gamma_e$ in the comoving frame, the synchrotron cooling rate is given by \citep{rybickiRadiativeProcessesAstrophysics1979a}
\begin{equation}
\dot{\gamma}_{e,\rm{SYN}}=-\frac{\sigma_{T} B^{2} \gamma_{e}^{2}}{6 \pi m_{e} c}.
\end{equation}
Here, $\sigma_T$ is the Thomson cross section, $m_e$ is the electron mass, $c$ is the speed of light, and $B$ is the magnetic field, respectively. The adiabatic cooling rate can be described as (e.g., \citealt{moderskiBeamingEffectsAfterglow2000,tavecchioClumpsLargeScale2003,uhmDynamicsAfterglowLight2012})
\begin{equation}
\dot{\gamma}_{e,\rm{ADI}}=-\frac{2}{3} \frac{\gamma_{e}}{t},
\end{equation}
where $t$ is the comoving time and a two-dimensional expansion is considered. The SSC cooling rate is \citep{jonesCalculatedSpectrumInverseComptonScattered1968,boettcherGrayEmissionSpectral1997a}
\begin{equation}
\dot{\gamma}_{e,SSC}=-\frac{3 \sigma_T m_e c^3}{8 h^2 }
\int_{0}^{\infty} \frac{u'(\nu')}{\nu'^2} {\cal G}(\gamma_{e},\nu')d\nu'
\end{equation}
where ${\cal G}(\gamma_{e},\nu')$ are given by
${\cal G}(\gamma_{e},\nu')=\displaystyle\frac{8}{3}\frac{E(1+5E)}{(1+4E)^2}-\frac{4E}{1+4E}(\frac{2}{3}+\frac{1}{2E}+\frac{1}{8E^2})+\rm{ln}(1+4E)[1+\frac{3}{E}+\frac{3}{4}\frac{1}{E^2} 
\displaystyle+\frac{\rm{ln}(1+4E)}{2E}-\frac{\rm{ln}(4E)}{E}] -\frac{5}{2}\frac{1}{E}+\frac{1}{E}\sum_{n=1}^{\infty}
\frac{(1+4E)^{-n}}{n^2}-\frac{\pi^2}{6E}-2 $ for $E>1/4$, and ${\cal G}(\gamma_{e},\nu')=\displaystyle E^2 (\frac{32}{9}-\frac{112}{5}E+\frac{3136}{25}E^2)$ otherwise. $E=\gamma_{e}h\nu'/m_{e}c^2$, $\nu'$ is the comoving frequency of seed photons and $h$ is the Planck constant. 

A numerical method is used to solve the kinetic equations of the electron distribution \citep{chiabergeRapidVariabilitySynchrotron1999}:
\begin{equation} \label{eq:electron} \frac{\partial}{\partial R} \left(\frac{{\rm d}N_{\rm e}}{{\rm d}\gamma_{\rm e}} \right) + \frac{\partial}{\partial \gamma_{\rm e}} \left[ \frac{{\rm d} \gamma_{\rm e}}{{\rm d} R} \left(\frac{{\rm d}N_{\rm e}}{{\rm d}\gamma_{\rm e}} \right) \right]=Q\left( \gamma_{\rm e} \right), 
\end{equation} 
where ${\rm d}N_{\rm e}/{\rm d}\gamma_{\rm e}$ is the electron distribution, $Q\left( \gamma_{\rm e}\right)$ is the electron injection rate with units of ${\rm cm}^{-1}$, $\frac{d\gamma_e}{dR}=(\dot{\gamma}_{e,\rm{SYN}}+\dot{\gamma}_{e,\rm{SSC}}+\dot{\gamma}_{e,\rm{ADI}})/c\beta_\gamma\gamma$, and $\beta_\gamma=\sqrt{\gamma^2-1}/\gamma$. In our emission calculation, the SSA and high energy absorption due to the photon pair production within the emission region are also taken into account numerically \citep{lithwickLowerLimitsLorentz2001,zhaoBulkLorentzFactors2011}.

\subsection{resulting numerical light curves}

\begin{figure}[htbp]
    \centering
    \includegraphics[width=1\textwidth]{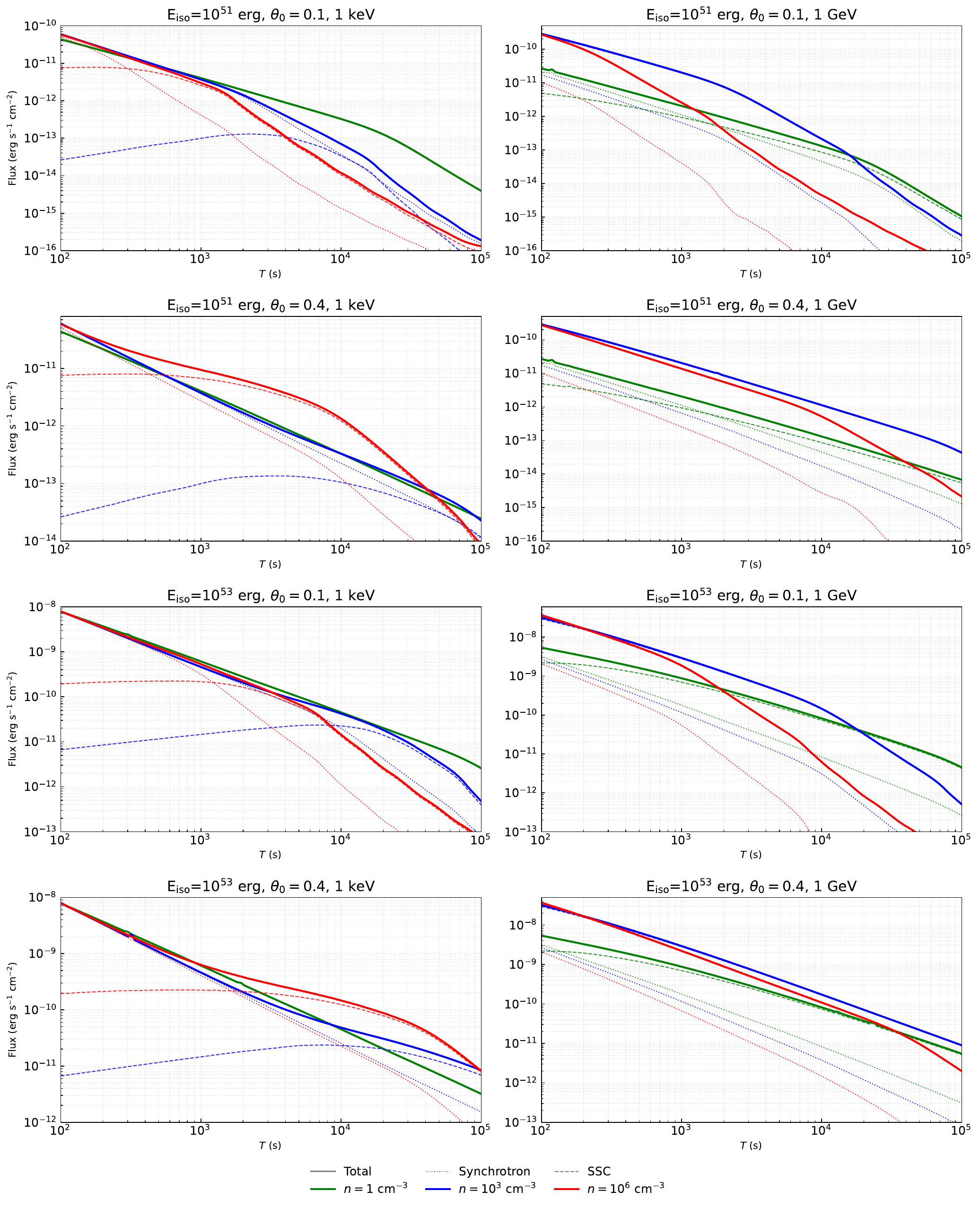} %
\caption{Comparison of light curves for different emission components (SSC, synchrotron, and total of the two components) in 1 keV and 1 GeV bands under varying jet half opening angle $\theta_0$, ambient density $n$, and the isotropic equivalent energy $E_{\rm iso}$. Fixed parameters:  $\epsilon_B = 0.001$, $\epsilon_e = 0.1$ $p=2.3$.}
    \label{fig:comp_ssc_syn_comp} %
\end{figure}

We calculate multiple sets of afterglow light curves using different physical parameters to explore the relative contributions of synchrotron and SSC emission components across three bands. The relative strength between the SSC  and synchrotron components depends on the ratio
$\epsilon_e/\epsilon_B$, with larger ratios resulting in a more significant SSC contribution. To illustrate the effect of the SSC component on the light curves, we adopt a large ratio $\epsilon_e/\epsilon_B=1000$ and present in Fig. \ref{fig:comp_ssc_syn_comp} the decomposition of the synchrotron and SSC components under different parameter sets. It can be clearly seen that the SSC component plays a crucial role in shaping the X-ray (keV band) light curves. In particular, it can give rise to a characteristic rising phase in the case of a large jet opening angle ($\theta_0=0.4$), which originates from the temporal behavior of the SSC emission, $F_\nu^{\rm SSC}\propto T^{1/8}$ (see eq.\ref{Fnussc_T}). When both synchrotron and SSC components are included, the total light curve often exhibits a shallow decay phase or even a bump-like feature (for very large ratio of $\epsilon_e/\epsilon_B$) before the jet break. This behavior originates from the emission component transition in the afterglow evolution from a phase initially dominated by synchrotron emission to a later phase increasingly governed by SSC emission. As a comparison, we also show in the figure the light curves for the typical ISM case with $n=1$ cm$^{-3}$. It is noted that, in this case, no distinct features appear in the X-ray band.

\begin{figure}[htbp]
    \centering
    \includegraphics[width=1\textwidth]{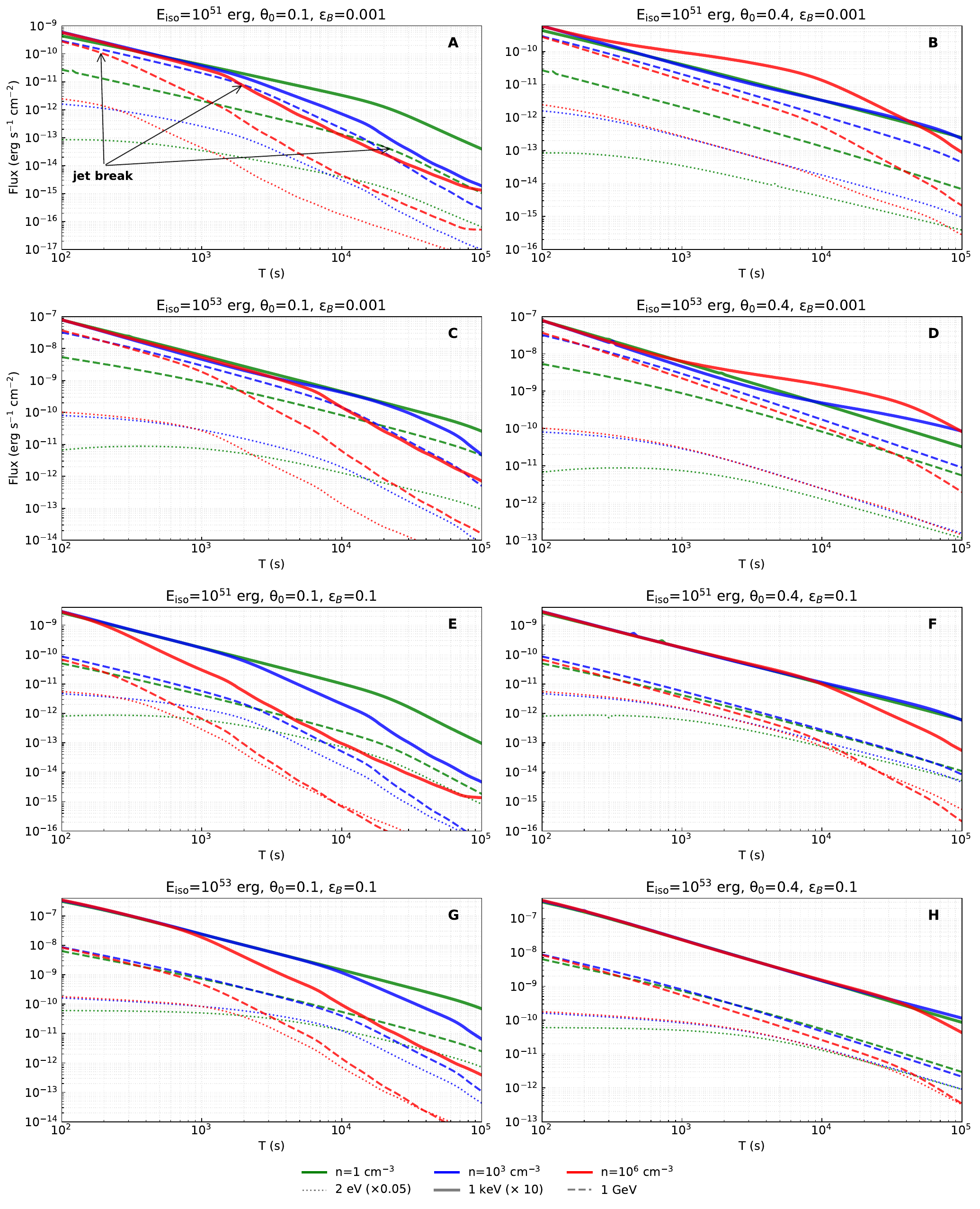} %
\caption{Comparison of total light curves (including the synchrotron and the SSC components) in 2 eV (the optical band), 1 keV, and 1 GeV under varying parameters. In panel A, the jet breaks in the GeV band (dashed lines) are explicitly marked. Fixed parameters are the same to Fig. \ref{fig:comp_ssc_syn_comp} .}
    \label{fig:comp_keV_GeV} %
\end{figure}

Figure~\ref{fig:comp_keV_GeV} shows the total light curves (including the synchrotron and SSC components) in the optical, X-ray (keV), and high-energy (GeV) bands for a direct comparison of multi-wavelength afterglow behavior. Two representative cases are considered: a large ratio case of $\epsilon_e/\epsilon_B=1000$ and a small ratio case of $\epsilon_e/\epsilon_B=1$. In the case of $\epsilon_e/\epsilon_B=1000$ (see panels A-D in Fig. \ref{fig:comp_keV_GeV}), the timing and prominence of this transition strongly depend on the ambient medium density n. In a typical ISM environment, the early afterglow is dominated by synchrotron emission, while SSC gradually becomes more important at later times. However, the contributions from both components remain broadly comparable throughout the evolution. In contrast, in a denser external medium, the SSC component becomes the dominant emission mechanism from the very beginning. Moreover, the higher the ambient density, the stronger the SSC emission becomes relative to the synchrotron component, thereby playing an increasingly dominant role in shaping the observed light curves. 

In contrast, such shallow decay features in the dense medium are not observed in the high-energy (e.g., GeV) and optical bands. These distinct features arise from the differing dominance of emission mechanisms in these bands. In the GeV band, the SSC component remains dominant from very early afterglow, leading to a smoothly decaying light curve that lacks the transitional features seen in the X-ray band. On the other hand, in the optical band, the emission is governed by synchrotron radiation alone throughout the afterglow phase, resulting in typical afterglow light curves that show no significant contribution from SSC emission.

From Fig. \ref{fig:comp_ssc_syn_comp} and \ref{fig:comp_keV_GeV}, we can also find that in denser environments, the shallow decay phase in the X-ray band emerges earlier when the jet has a large opening angle. By comparison, for a more narrow jet (e.g., $\theta_0=0.1$), the jet break occurs at an earlier time, $T_j \approx 104 E_{\rm iso,51}^{1/3} n_{6}^{-1/3} \Big(\frac{\theta_0}{0.1}\Big)^{8/3} $s, and we do not observe a distinct X-ray shallow decay or bump. However, the jet break displays wavelength-dependent behavior due to the different dominance of emission mechanisms. The GeV and optical light curves show a simultaneous jet break. This simultaneity is consistent with theoretical models where the jet edge becomes visible to the observer and the lateral expansion becomes significant. Such consistency is expected, as the emission in the GeV and optical bands is dominated by a single mechanism, namely the SSC and synchrotron processes, respectively. On the other hand, in the keV band, the transition from synchrotron to SSC dominance compensates for the post-jet-break decline, causing the jet break to be delayed relative to its timing in the GeV and optical bands. 
This temporal offset between the jet breaks in the three bands could represent a potential signature of high-density environments.

In the case of $\epsilon_e/\epsilon_B=1$ (see panels E–H in Fig.~\ref{fig:comp_keV_GeV}), the SSC component is not prominent, and the synchrotron emission dominates across all three bands. As a result, the light curves in the optical, X-ray, and GeV bands appear similar. The only noticeable difference appears in the early-time optical band, where the light curve is initially flat due to the passage of $\nu_m$ through the optical band (see the evolution of $\nu_m$ in Fig. \ref{evolu_nu} and the corresponding light curve slopes in \citealt{1998ApJ...497L..17S}). After the crossing, the optical light curves are similar to those in the other bands.

As anticipated from the analytical estimates (Section \ref{analytical_lc} and Fig. \ref{lc_analy}), the GeV light curves in the numerical models, prior to the jet break for wide jet opening angles, exhibit a steady power-law decay dominated by SSC emission, while the X-ray band shows the predicted transition from synchrotron to SSC. The numerical results thus closely follow the analytical expectations, confirming that the simplified treatment captures the essential behavior. The main differences appear for narrower jets, where the earlier jet break modifies the timing and prominence of the SSC features. These effects are not included in the analytical treatment.

In summary, our numerical results across a comprehensive parameter space reveal several clear trends. (i) The relative importance of SSC versus synchrotron is primarily controlled by $\epsilon_e/\epsilon_B$, with large ratios yielding pronounced SSC-dominated phases, particularly in the X-ray band. This is expected since in dense environments electrons are typically in the fast-cooling regime, and thus the Compton $Y$ parameter scales with $\epsilon_e/\epsilon_B$. (ii) The ambient density $n$ regulates the timing and prominence of the SSC contribution: higher densities enhance SSC and shift the X-ray shallow decay to earlier times. (iii) The jet opening angle $\theta_0$ governs the emergence of X-ray bump-like features: wide jets postpone the jet break and allow the SSC transition to manifest, whereas narrow jets cause an earlier, frequency-dependent jet break. (iv) The GeV and optical bands are each dominated by a single mechanism (SSC and synchrotron, respectively), and thus their light curves evolve smoothly and exhibit simultaneous jet breaks.

\clearpage

\subsection{Comparison with Observations: GRB 191019A} \label{comp_ob_data}
\begin{figure}[htbp]
    \centering
    \includegraphics[width=1\textwidth]{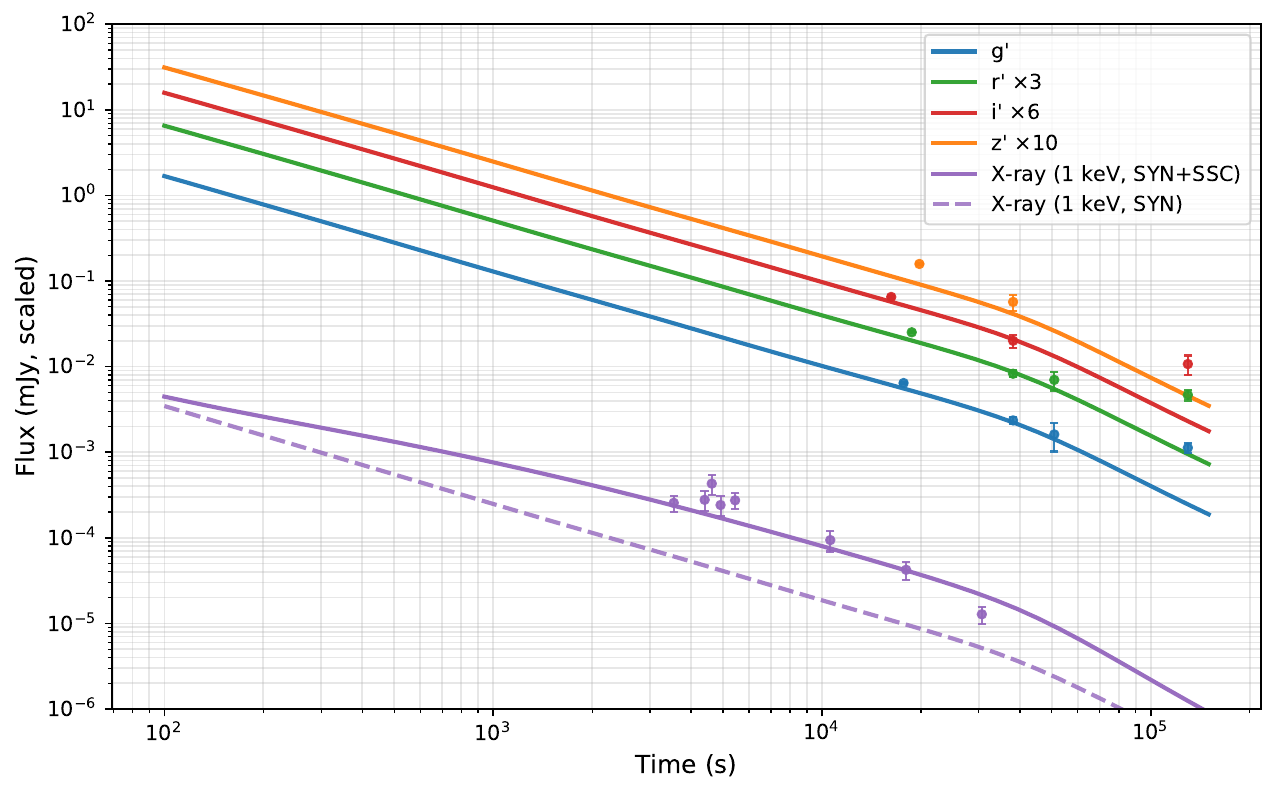} %
\caption{Light curves generated from our numerical modeling (solid lines) compared with the observed data (points) of GRB 191019A. 
The observational data are taken from \citet{2023NatAs...7..976L} and \citet{strattaPuzzlingLongGRB2025}. The purple solid line indicates the total light curve in the keV band, including both the synchrotron and SSC components, while the purple dashed line represents the synchrotron-only contribution. The parameters adopted in the modeling are 
$E_{\rm iso} = 2.9\times10^{51}\ \mathrm{erg}$, 
$\epsilon_B = 1.0\times10^{-5}$, 
$\epsilon_e = 2.8\times10^{-2}$, 
$n = 4.7\times10^{4}\ \mathrm{cm^{-3}}$, 
$\theta_0 = 0.45\ \mathrm{rad}$, 
and $p = 2.3$. 
}
    \label{fig:modeling_to_data} %
\end{figure}

To illustrate the applicability of the proposed scenario, we perform a case study of GRB~191019A.  
By exploring a broad range of physical parameters, we identify a set of conditions that reproduce the overall temporal behavior of the observed afterglow light curves in the X-ray and optical bands, under the assumption of a dense circumburst medium with $n \sim 5\times10^{4}~\mathrm{cm^{-3}}$ and moderate microphysical parameters (see Fig.~\ref{fig:modeling_to_data}).  
As shown in the figure, the light curve in the X-ray band is dominated by the SSC component. 
This demonstrates that a dense environment can provide parameter spaces consistent with the observational data.  
We also note that the flux in the $g'$, $r'$, and $i'$ bands at $\sim1.5$ days ($\sim 1.3\times10^5$ s) appears to be brighter than the model light curves, suggesting the presence of an additional emission component, possibly a kilonova as proposed by \citet{strattaPuzzlingLongGRB2025}.

\section{discussion and conclusion}

Our theoretical analysis shows that GRBs in dense circumburst environments can produce distinctive afterglow signatures, characterized by band-dependent light curve behaviors. Specifically, for large jet opening angles, a shallow decay or bump-like features may appear in the X-ray band, while such features are absent in the GeV and optical bands. Moreover, for small jet opening angles, the optical and GeV bands exhibit simultaneous jet breaks, whereas the X-ray band shows a delayed jet break. These distinctive behaviors can serve as indicators of a dense environment, such as AGN disks.

GRB 191019A is argued to be a burst that happened in an AGN disk \citep{lazzatiPromptEmissionGRay2022,lazzatiGRB191019AShort2023a,2023NatAs...7..976L}. 
The X-ray afterglow observed by Swift/XRT can be fitted with a single power-law decay, but a broken power-law model provides an alternative fit with temporal indices of approximately $\sim 0.14$ and $\sim -1.56$ (from UKSSDC\footnote{\url{http://www.swift.ac.uk/xrt\_curves/}}, \citealt{evansMethodsResultsAutomatic2009}). This may suggest the influence of a dense surrounding medium, with the shallow rise indicating the contribution of SSC emission. 
However, the poorly sampled X-ray light curves and the lack of well-sampled multi-wavelength observations limit the robustness of this interpretation. 
Although the current evidence remains inconclusive, when combined with other observational signatures of dense surroundings, such as prompt emission features \citep{lazzatiPromptEmissionGRay2022,lazzatiGRB191019AShort2023a} and occurring near the center of its host galaxy \citep{2023NatAs...7..976L}, the afterglow signature discussed here could provide additional support for the scenario that the burst occurred in a dense AGN disk.

Alternative interpretations of GRB 191019A have also been proposed. \citet{strattaPuzzlingLongGRB2025} modeled the available optical and X-ray afterglow data with a low-density circumburst medium and argued for a possible kilonova contribution. Given the limited sampling and the number of free parameters, their fit represents one viable scenario but does not strongly exclude alternatives. A dense environment therefore remains possible. 
Our modeling in Section~\ref{comp_ob_data} indicates that a dense-medium scenario can also account for the main features of GRB~191019A’s afterglow emission, suggesting that the burst possibly occurred in a dense environment. By contrast, \citet{wangEjectaCircumstellarMedium2024} investigated emission powered by the interaction between kilonova ejecta and a dense circumburst medium, showing that such interaction can give rise to a late-time emission component, which, however, could be concealed by the bright host galaxy in the case of GRB 191019A. These works indicate that the circumburst density of GRB 191019A remains inconclusive, as the present data do not provide sufficient constraints.

While our study focuses on GRBs in dense environments such as AGN accretion disks, our results are most relevant for cases where the burst originates in the relatively less dense outer regions of the disk, where the afterglow emission can emerge from the disk. In comparison, if a jet is launched deep inside a massive AGN disk, it may be choked before breakout 
(e.g., \citealp{2021ApJ...911L..19Z,2024ApJ...976...63Z}) and additional radiative transfer processes such as SSA heating (\citealp{ghiselliniThermalizationSynchrotronAbsorption1998,kobayashiCharacteristicDenseEnvironment2004,gaoComptonScatteringSelfabsorbed2013}) could significantly alter the afterglow emission. As shown in Fig. \ref{evolu_nu}, for the density up to $n = 10^9$ cm$^{-3}$, $\nu_a$ exceeds  $\nu_m$ at very early time and thus the SSA heating will become important \citep{gaoComptonScatteringSelfabsorbed2013}. 
Therefore, to avoid the complexities introduced by SSA heating, our current study limits the analysis to environments with densities up to $n=10^6$ cm$^{-3}$. A more detailed exploration of these high-density regimes will be presented in future works.

In the dense circumburst medium considered here, GeV photons are produced through SSC emission. As shown by eqs. \ref{Fnussc_T} and \ref{tausyn}, while the maximum photon energy is sensitive to the ambient density $n$, the GeV flux depends only weakly on $n$, as also illustrated in Figs. \ref{fig:comp_ssc_syn_comp} and \ref{fig:comp_keV_GeV} prior to the jet break. This indicates that dense media by themselves do not significantly alter the flux level compared to more typical densities. Nevertheless, higher densities also lead to an earlier jet break, after which the afterglow emission enters a rapid decay phase, with the GeV afterglow in particular becoming difficult to detect at late times. Thus, in the case of narrower jets, useful GeV signals are expected predominantly at earlier times, where the jet break can still be captured. When combined with multi-wavelength observations, the temporal behavior of GeV emission therefore provides a valuable diagnostic of the circumburst density. We do not include external inverse Compton processes with disk photons as seeds, nor the possible attenuation of GeV photons by the disk radiation field. However, in the outer regions of AGN disks, where the disk photon density is relatively low, both effects are expected to be negligible. \citet{yuanGeVSignaturesShort2022} have shown that even at outer disk densities up to $\sim 10^{14}\mathrm{cm^{-3}}$, disk photons do not strongly suppress GeV emission. Therefore, within the parameter space explored in this work ($n \lesssim 10^{6}\mathrm{cm^{-3}}$), dense circumburst environments are not expected to hinder the production or escape of GeV photons.

In the early afterglow phase, approximately one-third of X-ray afterglows exhibit a so-called shallow decay phase that lasts up to $10^{4\text{--}5}$ seconds. \citep{zhangPhysicalProcessesShaping2006,2007ApJ...670..565L,2019ApJ...883...97Z}. This behavior resembles that expected when a GRB occurs in a dense environment. However, this shallow decay is generally believed to be due to the energy injection \citep{daiGammarayBurstAfterglows1998a,zhangGammaRayBurstAfterglow2001,zhangPhysicalProcessesShaping2006}, and the spectral index typically remains constant across both the shallow and the subsequent normal decay phases \citep{2007ApJ...670..565L,2019ApJ...883...97Z}. In contrast, if the shallow decay is caused by an increasing SSC contribution in a dense circumburst environment, the spectral shape is expected to evolve with time, as the dominant radiation mechanism transitions from synchrotron to SSC and the emission regime changes from the shallow decay to the follow-up decay. This spectral evolution, combined with the multi-wavelength temporal behavior, provides a potential observational diagnostic method to distinguish between these scenarios.

As discussed above, the strength of the SSC component is highly sensitive to the ratio $\epsilon_e/\epsilon_B$. Theoretically and observationally, $\epsilon_e$ is typically found to be around 0.1, while $\epsilon_B$ remains highly uncertain, with values ranging from $\sim 0.1$ down to as low as $10^{-7}$ in different GRBs \citep{medvedevGenerationMagneticFields1999,2014ApJ...785...29S}. This wide range of $\epsilon_B$ implies that the ratio $\epsilon_e/\epsilon_B$ can vary over several orders of magnitude. 
Thus, we expect that in a significant fraction of GRBs occurring in dense environments, the SSC component will play an important role in shaping the X-ray afterglow emission.

\begin{acknowledgments}
The author thanks the referee for constructive comments and suggestions. This study is supported by the National Natural Science Foundation of China (grants No. 12393811, 12473048).
\end{acknowledgments}

\bibliography{sample701}{}
\bibliographystyle{aasjournalv7}

\end{document}